\documentclass[preprint, 5p,times, twocolumn,12pts]{elsarticle}
\usepackage[T1]{fontenc}
\usepackage{textcomp}
\usepackage{amsmath}
\usepackage{amssymb}
\usepackage{graphicx}
\usepackage[francais]{babel}
\usepackage[utf8]{inputenc}
\usepackage[T1]{fontenc}
\usepackage{natbib, hyperref}


\usepackage{babel}
\begin{document}

\journal{NIM-A}

\title{Investigation of the dynamics of ionization induced injected electrons under the influence of beam loading effects}
\author[LPGP]{P. Lee\corref{cor1}}
\ead[cor1]{patrick.lee@u-psud.fr}
\author[LPGP]{T. L. Audet}
\author[LBNL]{R. Lehe}

\address[LPGP]{LPGP, CNRS, Univ. Paris-Sud, Universit\'e Paris-Saclay, 91405, Orsay, France}

\author[LBNL]{J.-L.Vay}
\cortext[cor1]{Corresponding author}
\address[LBNL]{Lawrence Berkeley National Laboratory, Berkeley, CA 94720, USA}
\author[LPGP]{G. Maynard}
\author[LPGP]{B. Cros\corref{cor1}}

\ead[cor1]{brigitte.cros@u-psud.fr}
\date{\today}
\begin{abstract}
In laser-driven wakefield, ionization induced injection is an efficient way to inject electrons in the plasma wave.  A detailed study on the beam dynamics under the influence of beam loading effects, which can be controlled by the concentration of nitrogen impurity introduced in the hydrogen gas was conducted. For a specific value of this percentage, the final energy of the high-energy electron bunch becomes nearly independent of the trapped positions, thus leading to a small energy dispersion. We also show that the final beam emittance is mainly determined by the injection process.
\end{abstract}


\maketitle

\section{\label{sec:level1}Introduction}
Laser-driven plasma waves are capable of sustaining orders of magnitude increases in accelerating gradient \cite{tajima_laser_1979,esarey_physics_2009,malka_laser_2012}. In experiments, acceleration gradients $>100\,\mathrm{GV/m}$ \cite{leemans_gev_2006,nakamura_gev_2007} have been demonstrated, making laser wakefield acceleration (LWFA) a promising way towards more compact high-energy accelerators with a wide range of applications.

Acceleration of an electron beam in a single laser plasma accelerator stage is limited to a length determined either by diffraction, depletion of laser driver, or dephasing of electrons. Multistage acceleration schemes, consisting of an injector, a transport line and an accelerator \cite{leemans_laser-driven_2009}, are considered as one of the solutions to the next energy frontier. In these schemes, the electron injector is expected to produce a high-quality electron beam with narrow energy spread and small emittance. Many efforts have since been devoted to the control of the electron beam properties in the injector \cite{geddes_high-quality_2004,faure_controlled_2006,mangles_monoenergetic_2004}.


Ionization induced injection scheme \cite{mcguffey_ionization_2010,pak_injection_2010,clayton_self-guided_2010}, with the use of trace atoms brings about an additional degree of freedom, allows electron trapping at lower plasma densities, and use of lower laser intensities as compared to the self-injection scheme \cite{mangles_monoenergetic_2004,geddes_high-quality_2004,faure_laserplasma_2004,kalmykov_electron_2011}. However, the major disadvantage of the ionization induced injection scheme is the large energy spread of generated electrons due to continuous injection, as long as no competing mechanisms such as beam loading effects or the laser intensity decreases below the injection threshold are in play. Several methods to reduce the energy spread have been proposed \cite{liu_all-optical_2011,pollock_demonstration_2011,gonsalves_tunable_2011,kim_enhancement_2013,wang_quasi-monoenergetic_2013,golovin_tunable_2015} throughout the years. These methods use a few mm-long mixed gas volume followed by a volume where pure gas is injected, the second volume acting as an accelerator and energy filter. In the same line of thought, a detailed investigation on utilizing beam loading effects to reduce energy spread of the accelerated electron bunch was carried out \cite{lee_optimization_2017}.

This article provides complementary results to \cite{lee_optimization_2017}, in which the ionization induced injection scheme was studied for a mixture of hydrogen and nitrogen. In this scheme, the hygrogen gas is used to control precisely the background plasma electron density, whereas the nitrogen gas is used to tune independantly the number of injected electrons. We report on the trajectory analysis by analyzing the correlation between the injection position and the final properties of the high-energy electrons. We demonstrate that, at an optimum nitrogen concentration, beam loading effects lead to a final energy which is nearly independent on the injection position, yielding to the creation of a sharp peak distribution in the energy spectrum of the accelerated electron bunch. Regarding the emittance, our results also show that, the larger value observed in the laser polarization direction is mainly due to  the injection process.

The remaining of the paper is organized as follows. In Section \ref{sec:numerical} are reported the laser-plasma parameters and the numerical setup for PIC simulations. The trajectory analysis of the trapped electrons is discussed in Section \ref{sec:analysis}. Finally, the evolution of the beam emittance is presented in Section \ref{sec:second-order}.

\section{Choice of laser-plasma parameters}
\label{sec:numerical}
\subsection{Regime of acceleration}
\begin{figure}[htb]
\includegraphics[scale=0.95]{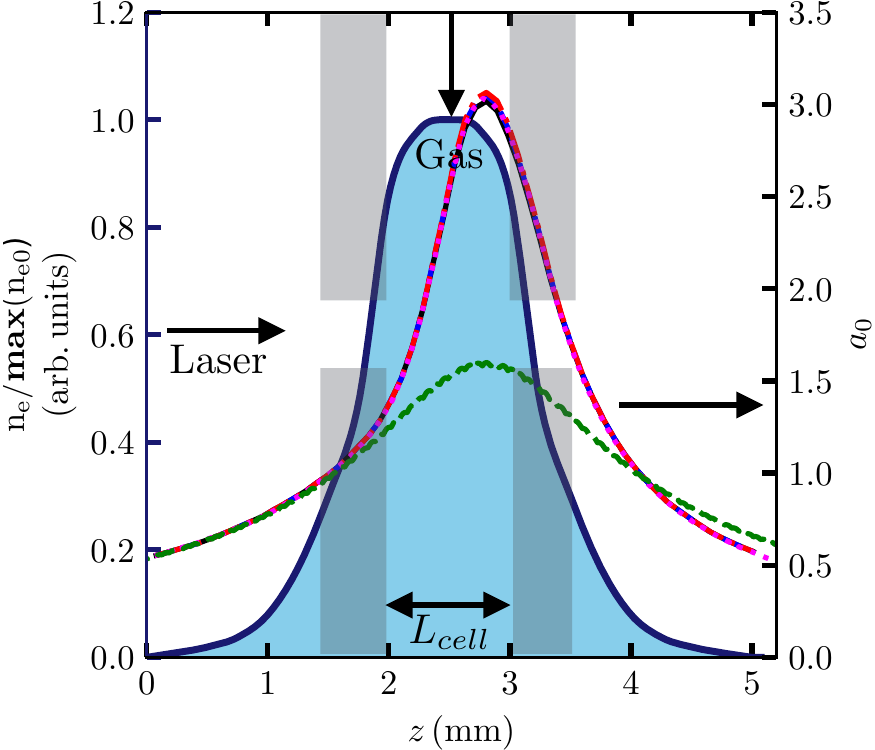}

\caption{The light blue area indicates the longitudinal density profile
of the gas cell (left vertical axis). Evolution of $a_{0}(z)$ with respect to the propagation axis $z$ (right vertical axis) for propagation in vacuum (green dashed line) and in plasma at different $\mathrm{C_{N_2}}$: $2\%$ (black solid line), 1\% (blue dashed line), 0.5 \% (red dashed-dotted line) and 0.35 \% (magenta dotted line). The gas cell is equipped with a gas inlet at the top and adjustable plates at the entrance and the exit used to modify the density profile from both ends, these plates of $500\,\mathrm{\mu m}$ in our study configuration are represented by the gray area. The length of the gas cell, $L_{cell}=1\,\mathrm{mm}$. The laser propagates from left to right. \label{fig:fig1}}
\end{figure}

An in-house gas cell, known as ELISA \cite{audet_electron_2016}, which stands for ELectron Injector for compact Staged high-energy Accelerator, was used to confine hydrogen gas, and a small fraction of trace atoms (typically nitrogen) for the ionization induced injection scheme experiments. ELISA allows for a modification of the density profile with adjustable parameters. The present numerical study is conducted in a specific configuration that was characterized experimentally and by using dynamic fluid simulations with the SonicFoam solver in openFOAM \cite{weller_tensorial_1998}, with its longitudinal density profile shown in Fig.~\ref{fig:fig1} \cite{audet_gas_nodate}.

Several criteria are taken into consideration in making the choice of the  laser-plasma parameters in this study. The injector should deliver an electron beam with an energy range between $\mathrm{50-200\,MeV}$. The lower limit is fixed at $\mathrm{50 \, MeV}$ to avoid the dominance of space charge effects, and to minimize the energy spread as it scales as $1/\gamma^{2}$, where $\gamma = (1-(v/c)^2)^{-1/2}$, is the Lorentz factor, $v$ the velocity of the electron and $c$, the speed of light. On the other hand, the upper limit is fixed at $\mathrm{200\,MeV}$ to allow for a compact transport line for electron beam manipulation before coupling to the first accelerating structure. In this study, we have chosen to accelerate electrons up to $150\,\mathrm{MeV}$. The required normalized transverse emittance of the electron bunch has to be small, $\varepsilon_n\sim 1\,\mathrm{mm\,mrad}$, whereas the energy spread should be $<10\%$ and the charge should be above $\mathrm{\geq}$ 10 $\mathrm{pC}$.

The optimization work has the objective to  produce a quasi-monoenergetic beam with the maximum of charge in the energy range of $150\,\mathrm{MeV}$. It relies on the optimization of the phase space rotation by choosing an acceleration length $L_{acc}$ close to the electron dephasing length, such that $L_{acc}\propto (\lambda_p^3/\lambda_0^2)a_0\propto \max (n_{e0})^{-3/2}$, with $\lambda_p$ plasma wavelength, $\lambda_0$ laser wavelength,  $a_0$ the vector potential of the laser pulse propagating in the vacuum at the focal point, and $n_{e0}$ the maximum electron number density on axis. Considering all these factors and results from the previous studies \cite{lee_optimization_2017, lee_dynamics_2016}, the plasma density is fixed at  $\max (n_{e0})$ to $4\times10^{18}\mathrm{cm}^{-3}$, with a gas cell length of $\sim1 \mathrm{mm}$.

The maximum value of the laser amplitude in normalized units is defined by $a_0(z) = \max_{r,t}[ea(r,z,t)/m_e\omega c]$, where $\omega$ is the laser frequency, $e$ the electron charge, and $m_e$ the electron mass. $a_0$ of the laser pulse is chosen to be $1.6$ as it is large enough to ionize and inject the $6^{th}$ electron of nitrogen but not too large to provoke self-trapping of electrons after self-focusing in the plasma, the limit of which is $\sim 4$ \cite{pak_injection_2010}. This value of $a_{0}(z)=1.6$ corresponds to the maximum value of laser amplitude at the focal plane longitudinal position in vacuum, $z=z_{f}$. In our simulations, the laser pulse is assumed to have Gaussian temporal and spatial profiles, with a laser duration, $\tau_L= 20\,\mathrm{fs}$ at full-width at half-maximum (FWHM), and a laser waist, $w_L= 1\mathrm{\mu m}$ at $1/e^2$ of the laser intensity, for efficient excitation of the plasma wave. The laser is also focused at the exit of the gas cell so as to get an increase of the plasma longitudinal field after the injection process in order to reduce the energy spread as will be demonstrated in next sections.

For this study, we have performed simulations for several nitrogen concentrations, $C_\mathrm{{N_2}}$ ranging from $0.35\%$, $0.5\%$, $1\%$, and $2\%$. Electrons created during the ionization of hydrogen and L-shell (outer shell) of nitrogen early at the front of the laser pulse, move collectively under the action of the ponderomotive force, resulting in the plasma wave structure behind the laser pulse. With the parameters considered in this study, K-shell (inner shell) electrons of nitrogen are ionized only close to the peak of the laser pulse; they can thus be injected at a phase of the plasma wave favoring their local trapping into the existing plasma structure. In order to keep the density profile independent of $C_\mathrm{{N_2}}$, the total density, $n_e(z)$ with $z$ the coordinate along the laser axis in the gas cell, is adjusted according to the relation $n_{e}(z) =n_{at}(z)[1+4C_\mathrm{N_2}]$, with $n_{at}$ total density of atoms.

\subsection{Numerical setup}

Simulations were performed with WARP \cite{vay_novel_2012} using the azimuthal Fourier decomposition algorithm \cite{lifschitz_particle--cell_2009,davidson_implementation_2015,lee_modeling_2015} in cylindrical geometry, and a field ionization module \cite{desforges_dynamics_2014} to describe ionization modules based on the ADK model \cite{ammosov_tunnel_1986}.

The mesh resolution was chosen to be $\Delta z =\lambda_0/25$ and $\Delta r= \lambda_0/6$ in the longitudinal and transverse directions.

Note that the simulation was performed up to few hundreds of $\mu m$ away from the exit of the gas cell, at positions where the plasma wakefield is nearly zero. At these positions, the divergence of the electron beam has already reduced the space charge force, therefore electrons propagate nearly without interaction.

\section{Influence of nitrogen concentration on trapped electron dynamics}
\label{sec:analysis}
\subsection{Laser plasma interaction}

In Fig.~\ref{fig:fig1} is illustrated the ELISA longitudinal density profile (left vertical axis) and the evolution of $a_0(z)$ while interacting with the plasma for all $C_\mathrm{N_2}$ and in green dashed curve the evolution of $a_0(z)$ of a laser propagating in vacuum, with $z$ the laser axis (right vertical axis). In vacuum, the $a_0$ value attains $1.6$ at $z_f=2.8\,\mathrm{mm}$, as expected. When the plasma was introduced, the laser self-focused and increased the value of $\max(a_{0}(z))$ by nearly a factor of 2, compared to the vacuum case. We also observe that the evolution of $a_0(z)$ is independent of $C_\mathrm{N_2}$, implying that the generated plasma waves are similar in all cases. Any further modification to the plasma waves during injection and acceleration processes can only be caused by the beam loading effects, which are directly controlled by $C_\mathrm{{N_2}}$.

When keeping $n_{e0}=\max\left[n_{e}\left(z\right)\right]$ constant, the influence of $C_\mathrm{{N_2}}$ on the laser propagation and the plasma wake generation is insignificant. Ionization and trapping of electrons coming from the ionization process $\mathrm{N}^{5+}\rightarrow \mathrm{N}^{6+}$ take place efficiently for $a_{0}>1.5$, occurring at $z=2.1-2.2\, \mathrm{mm}$, close to the density plateau but still at the density up ramp. $200\,\mathrm{\mu m}$ further, at $z=2.3\,\mathrm{mm}$, $\mathrm{N}^{6+}\rightarrow \mathrm{N}^{7+}$ ionization process begins, requiring an $a_{0}\geq2.0$. For the rest of the article, we denote the $6^{th}$ electron the electron created from the ionization process $\mathrm{N^{5+}}\rightarrow\mathrm{N^{6+}}$ and the $7^{th}$ electron from $\mathrm{N^{6+}}\rightarrow\mathrm{N^{7+}}$.

\begin{table}[htb]
\centering
\caption{Electron injection starting, end positions, and length for each $C_\mathrm{N_2}$.}
\label{tab:tab2}
\begin{tabular}{c|c|c|c}
\begin{tabular}[c]{@{}c@{}}$C_\mathrm{N_2}$ \\ ($\%$)\end{tabular} & \begin{tabular}[c]{@{}c@{}}Starting position\\ $z\,\mathrm{(mm)}$\end{tabular} & \begin{tabular}[c]{@{}c@{}}End position\\ $z\,\mathrm{(mm)}$\end{tabular} & \begin{tabular}[c]{@{}c@{}}Length\\ $z\,\mathrm{(mm)}$\end{tabular} \\ \hline
0.35 & 2.24 & 3.21 & 0.97 \\
0.5 & 2.18 & 3.14 & 0.96 \\
1.0 & 2.09 & 2.97 & 0.88 \\
2.0 & 2.09 & 2.88 & 0.79
\end{tabular}
\end{table}

In Table.~\ref{tab:tab2} is reported the beginning, end positions, and the length of the injection process for each $C_\mathrm{{N_2}}$. We observe that the higher the $C_\mathrm{N_2}$, the earlier the injection process begins, and the shorter the injection length. The earlier injection position for high $C_\mathrm{{N_2}}$ is due to an abundance of nitrogen atoms, therefore the probability of getting injected into wakefields is higher at the earlier stage, whereas the shorter length can be explained by the truncation due to beam loading effects \cite{lee_optimization_2017}.

\subsection{Dynamics of trapped electrons}

\begin{figure}[htb]
\includegraphics{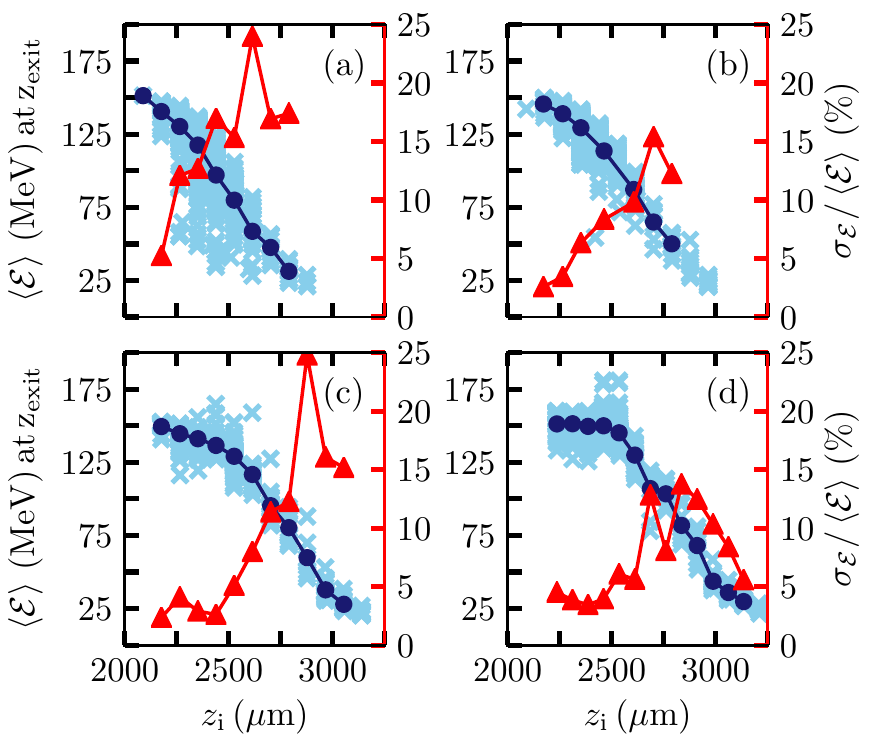}

\caption{Average energy at the exit of the simulation with respect to injected position, $z_\mathrm{i}$ for the $6^{th}$ electrons, for nitrogen concentration: (a) $2\%$, (b) $1\%$, (c) $0.5\%$ and (d) $0.35\%$. The average of average energy and energy dispersion per slice evaluated along $z_\mathrm{i}$ are represented respectively by the blue dots and the red triangles.\label{fig:fig2}}
\end{figure}

\begin{figure}[!htb]
\includegraphics{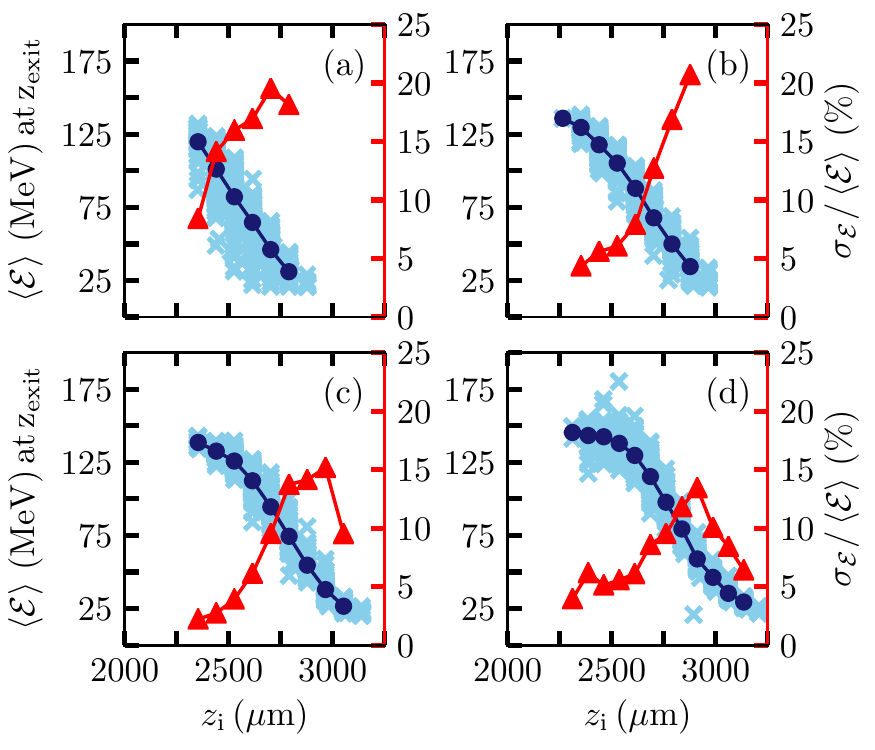}
\caption{Same figure as Fig.~\ref{fig:fig2} for the $7^{th}$ electrons. \label{fig:fig3}}
\end{figure}

In order to analyze the dynamics of trapped electrons, we backtracked 2000 randomly sampled trapped electrons (1000 for the $6^{th}$ and 1000 for the $7^{th}$ electrons) starting from $z_\mathrm{exit}=5.1\,\mathrm{mm}$ all the way back to their injected position, also corresponding to the position of their first appearance in the simulation. A study of correlation between the electron average energy at $z_\mathrm{exit}$ and its injection position, $z_\mathrm{i}$ are performed on the $6^{th}$, and the $7^{th}$ in Fig.~\ref{fig:fig2}, and in Fig.~\ref{fig:fig3} respectively. The blue dots and the red triangles represent respectively the average value of the average energy and the energy dispersion per slice along the injected position.

From Fig.~\ref{fig:fig2}(a-b), the average values of $\left<\mathcal{E}\right>$ represented by the blue dots show that earlier injected electrons gain a higher energy as compared to the later ones. This observation is due to the fact that earlier injected electrons are accelerated to a larger distance, as the energy gain is given by $\Delta\mathcal{E} = eL_{acc}E_z $, with $e$ the elementary charge, $L_{acc}$ the acceleration length and $E_z$ the average accelerating wakefield. In contrast, in Fig.~\ref{fig:fig2}(c-d), two populations can be identified. The first population suggests that electrons injected at $z_\mathrm{i}<\mathrm{2700\,\mu m}$ have a narrower average energy range at $z_{exit}$, between $125-180\,\mathrm{MeV}$, providing a peaked distribution in the energy spectrum, and these electrons constitute the high-energy electron bunch; the second population electrons injected at $z_i>\mathrm{2700\,\mu m}$, behave the same way as the electrons in Fig.~\ref{fig:fig2}(a-b). This behavior is related to beam loading effects. The lowest energy spread is obtained when $\Delta\mathcal{E}$ becomes independent of $L_{acc}$, implying that $E_z$, should increase with $z_i$ as $1/z_i$. With the laser being focused at the exit of the gas cell, the longitudinal plasma wakefield satisfies the aforementioned condition, as observed in Fig.~\ref{fig:fig2}(c-d) where beam loading effects are not prominent. However with a higher charge being trapped in the case of high $C_\mathrm{N_2}$, the accelerated electron bunch produces a wake that drastically modifies the fields of the accelerating wakefield, resulting in a decrease of the accelerating wakefield for electrons that are trapped at later times. As these electrons experience a lower accelerating wakefield as compared to the previously injected ones, their energy does not reach energy values equivalent to the high-energy electron bunch.

In terms of energy dispersion,  $\sigma_\mathcal{E}/\left<\mathcal{E}\right>$ is shown to increase with the injected position $z_\mathrm{i}$. In the case of $C_\mathrm{N_2}<1\%$, the earlier injected electrons with $z_\mathrm{i}<2700\,\mathrm{\mu m}$, have a high average energy $\left<\mathcal{E}\right>>125\,\mathrm{MeV}$, and a low energy dispersion $\sigma_\mathcal{E}/\left<\mathcal{E}\right>\leq5\%$. In fact, the energy dispersion per slice is given by $\sigma_{\mathcal{E}}(z) = \left<P(z)\Delta E_z(z)\right>$, where $P(z)$ and $\Delta E_z(z)$ are respectively the number of electrons and the gradient of the accelerating field over a length $\Delta z$.

In Fig.~\ref{fig:fig3}, the same tendency is retrieved for the $7^{th}$ electrons, only that they are trapped starting from $z_\mathrm{i}\sim 2300\,\mathrm{\mu m}$, which is $\sim 200 \,\mathrm{\mu m}$ further than the $6^{th}$ electrons. This $200\,\mathrm{\mu m}$ difference in the injection starting position leads to the $7^{th}$ electrons contributing mostly to the low energy end.


\section{Evolution of the transverse emittance of the accelerated electron bunch}
\label{sec:second-order}

\begin{figure}[htb]
\includegraphics{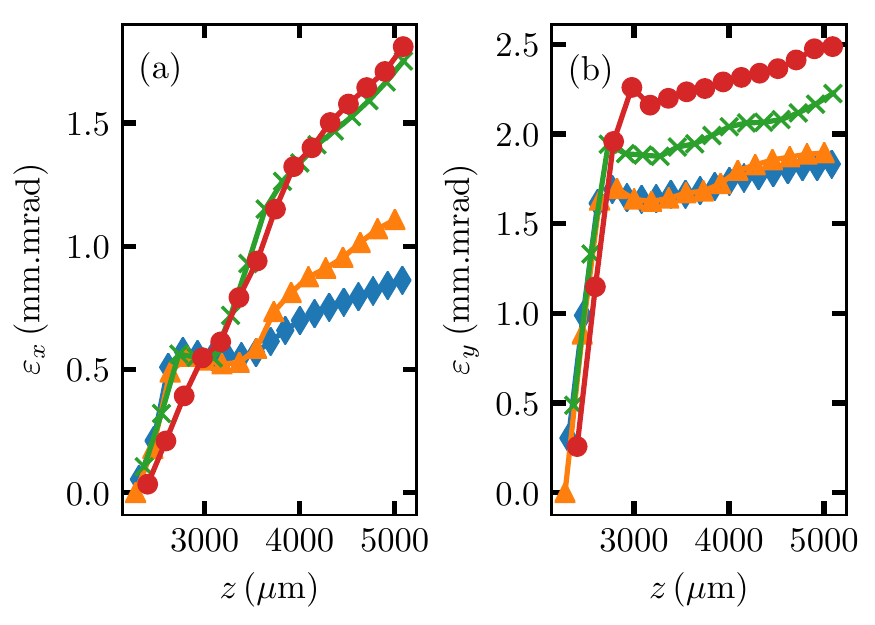}
\caption{Evolution of emittances (a) $\varepsilon_x$, and (b) $\varepsilon_y$ with respect to $z$. $\varepsilon_x$, and $\varepsilon_y$ are represented for the following concentrations: $2\%$ (in red dots), $1\%$ (in green crosses), $0.5\%$ (in orange triangles), and $0.35\%$ (in blue diamonds). \label{fig:fig4}}
\end{figure}

The beam emittance is a key parameter in designing the transport system to the accelerating stage of the multistage accelerator. The evaluation of the beam emittance presented here only takes into account electrons falling inside the distribution fitted by Gaussian after thresholding the beam spectrum profile at 50 \% of the peak value. This definition excludes contributions of low-energy electrons which are not relevant for injection into the second stage, and can dominate the rms beam emittance. Fig.~\ref{fig:fig4} shows the evolution of emittance in the transverse plane ($x,y$), with $y$ being the laser polarization axis, for all $C_\mathrm{N_2}$.

The normalized rms emittance is calculated using the standard formula $\varepsilon_{i,n}^2=\langle x_i^{2}\rangle \langle p_{i}/m_ec\rangle ^{2}-\langle x_ip_{i}/m_ec\rangle ^{2}$, where $x_i$ and $p_i$ are the electron position and momentum along the $i$-axis. As observed in Fig.~\ref{fig:fig4}(a), for all $C_\mathrm{N_2}$, there is first an increase to a value of $\sim0.5\,\mathrm{mm.mrad}$, then a plateau is observed beginning from $z\sim 2500\,\mathrm{\mu m}$, a signature of the end of injection of high-energy electrons, which then enter the acceleration phase. The length of this plateau correlates to the acceleration length, it is shorter in the case of high $C_\mathrm{N_2}$ because of beam loading effects. The further increase of emittance for $z>3000\,\mathrm{\mu m}$, situated in the down ramp of the plasma density profile, can be due to non-adiabatic evolution of the plasma wave or numerical inaccuracy, a further investigation is required to pinpoint the main cause.

In Fig.~\ref{fig:fig4}(b), the evolution of $\varepsilon_y$ shows a steep increase in the injection phase, at $z\sim 2500\,\mathrm{\mu m}$, the attained value is $>3$ times its counterpart in the $x-$direction, due to the laser polarization effect \cite{barber_measured_2017}, where a residual transverse momentum $p_\perp/m_ec\approx a(z_i)$ contributes to the emittance growth during the ionization process. Likewise, a slight inflection of the value is observed before the increase till the end of the simulation.

\begin{figure}[htb]
\includegraphics{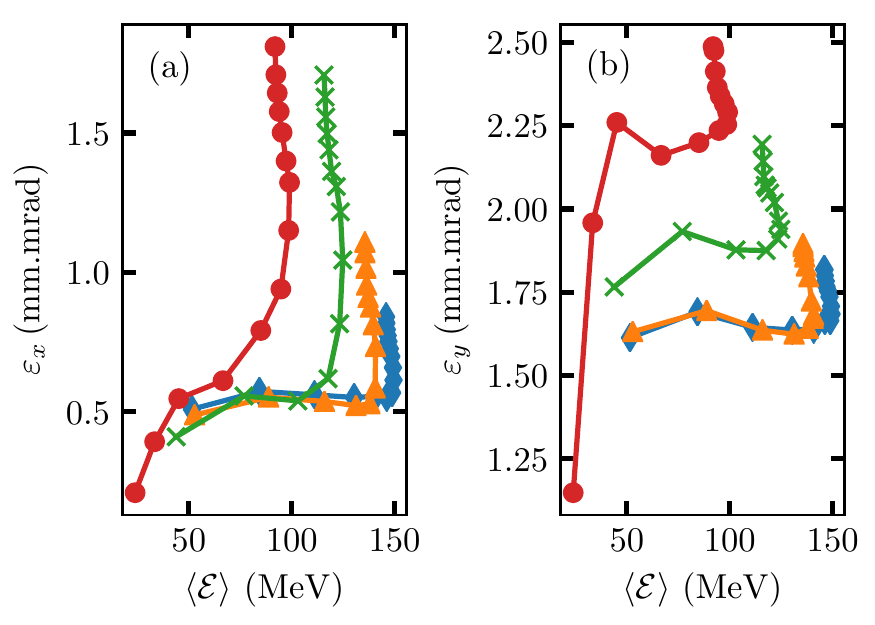}
\caption{Emittance (a) $\varepsilon_x$ and (b) $\varepsilon_y$ distribution with respect to average energy $\left<\mathcal{E}\right>$, at longitudinal positions, plotted for the nitrogen concentrations: $2\%$ (in red dots), $1\%$ (in green crosses), $0.5\%$ (in orange triangles), and $0.35\%$ (in blue diamonds).\label{fig:fig6}}
\end{figure}

Fig.~\ref{fig:fig6} shows the variation of emmitances (a) $\varepsilon_x$ and (b) $\varepsilon_y$ with respect to average energy, $\left<\mathcal{E}\right>$, for all $C_\mathrm{N_2}$. In Fig.~\ref{fig:fig6}(a), we observe a plateau at $\varepsilon_x=0.5\,\mathrm{mm.mrad}$, this plateau extends to certain values of average energy, where a steep increase is then observed. This suggests that $\varepsilon_x$ remains constant during the acceleration phase, and its growth only takes off when the acceleration is over, in the down ramp of the longitudinal plasma density profile. As for $\varepsilon_y$, as shown in Fig.~\ref{fig:fig6}(b), the initial plateau is higher than in $\varepsilon_x$, but a similar tendency is retrieved.






\section{Conclusion}
We have reported on a detailed analysis of the dynamics of the ionization induced injected electrons in a realistic laser-plasma configuration via the control of the nitrogen concentration. This analysis shows that for an optimized value of trace atom concentration, the final energy of the high-energy electron bunch becomes independent of their injection position, minimizing the energy spread. In addition, we have shown that the electron beam emittance is nearly independent of the accelerating phase, its growth is mainly during the injection and after the acceleration phases.

\section*{Acknowledgments}
We adknowledge the partial financial support of the Laboratoire d’Excellence PALM, within the Model\_LPA project. This work was granted access to the HPC resources of [TGCC/CINES] under the allocation 2017- [A0010510062] made by GENCI. We also acknowledge the use of the computing center MésoLUM of the LUMAT research federation (FR LUMAT 2764). T.L. Audet acknowledges financial support of EuPRAXIA, co-funded by the European Commission in its Horizon2020 Programme under the Grant Agreement no 653782.
\bibliographystyle{elsarticle-num}
\bibliography{ALP}

\begin{thebibliography}{10}
\expandafter\ifx\csname url\endcsname\relax
  \def\url#1{\texttt{#1}}\fi
\expandafter\ifx\csname urlprefix\endcsname\relax\def\urlprefix{URL }\fi
\expandafter\ifx\csname href\endcsname\relax
  \def\href#1#2{#2} \def\path#1{#1}\fi

\bibitem{tajima_laser_1979}
T.~Tajima, J.~M. Dawson,
  \href{http://link.aps.org/doi/10.1103/PhysRevLett.43.267}{Laser {Electron}
  {Accelerator}}, Physical Review Letters 43~(4) (1979) 267--270.
\newblock \href {http://dx.doi.org/10.1103/PhysRevLett.43.267}
  {\path{doi:10.1103/PhysRevLett.43.267}}.
\newline\urlprefix\url{http://link.aps.org/doi/10.1103/PhysRevLett.43.267}

\bibitem{esarey_physics_2009}
E.~Esarey, C.~B. Schroeder, W.~P. Leemans,
  \href{http://link.aps.org/doi/10.1103/RevModPhys.81.1229}{Physics of
  laser-driven plasma-based electron accelerators}, Reviews of Modern Physics
  81~(3) (2009) 1229--1285.
\newblock \href {http://dx.doi.org/10.1103/RevModPhys.81.1229}
  {\path{doi:10.1103/RevModPhys.81.1229}}.
\newline\urlprefix\url{http://link.aps.org/doi/10.1103/RevModPhys.81.1229}

\bibitem{malka_laser_2012}
V.~Malka, \href{http://dx.doi.org/10.1063/1.3695389}{Laser plasma
  accelerators}, Physics of Plasmas 19~(5) (2012) 055501.
\newblock \href {http://dx.doi.org/10.1063/1.3695389}
  {\path{doi:10.1063/1.3695389}}.
\newline\urlprefix\url{http://dx.doi.org/10.1063/1.3695389}

\bibitem{leemans_gev_2006}
W.~P. Leemans, B.~Nagler, A.~J. Gonsalves, C.~Tóth, K.~Nakamura, C.~G.~R.
  Geddes, E.~Esarey, C.~B. Schroeder, S.~M. Hooker,
  \href{https://www.nature.com/nphys/journal/v2/n10/full/nphys418.html}{{GeV}
  electron beams from a centimetre-scale accelerator}, Nature Physics 2~(10)
  (2006) 696--699.
\newblock \href {http://dx.doi.org/10.1038/nphys418}
  {\path{doi:10.1038/nphys418}}.
\newline\urlprefix\url{https://www.nature.com/nphys/journal/v2/n10/full/nphys418.html}

\bibitem{nakamura_gev_2007}
K.~Nakamura, B.~Nagler, C.~Tóth, C.~G.~R. Geddes, C.~B. Schroeder, E.~Esarey,
  W.~P. Leemans, A.~J. Gonsalves, S.~M. Hooker,
  \href{http://aip.scitation.org.proxy.scd.u-psud.fr/doi/abs/10.1063/1.2718524}{{GeV}
  electron beams from a centimeter-scale channel guided laser wakefield
  accelerator}, Physics of Plasmas 14~(5) (2007) 056708.
\newblock \href {http://dx.doi.org/10.1063/1.2718524}
  {\path{doi:10.1063/1.2718524}}.
\newline\urlprefix\url{http://aip.scitation.org.proxy.scd.u-psud.fr/doi/abs/10.1063/1.2718524}

\bibitem{leemans_laser-driven_2009}
W.~Leemans, E.~Esarey, \href{http://dx.doi.org/10.1063/1.3099645}{Laser-driven
  plasma-wave electron accelerators}, Phys. Today 62~(3) (2009) 44--49.
\newblock \href {http://dx.doi.org/10.1063/1.3099645}
  {\path{doi:10.1063/1.3099645}}.
\newline\urlprefix\url{http://dx.doi.org/10.1063/1.3099645}

\bibitem{geddes_high-quality_2004}
C.~G.~R. Geddes, C.~Toth, J.~v. Tilborg, E.~Esarey, C.~B. Schroeder,
  D.~Bruhwiler, C.~Nieter, J.~Cary, W.~P. Leemans,
  \href{http://dx.doi.org/10.1038/nature02900}{High-quality electron beams from
  a laser wakefield accelerator using plasma-channel guiding}, Nature
  431~(7008) (2004) 538--541.
\newblock \href {http://dx.doi.org/10.1038/nature02900}
  {\path{doi:10.1038/nature02900}}.
\newline\urlprefix\url{http://dx.doi.org/10.1038/nature02900}

\bibitem{faure_controlled_2006}
J.~Faure, C.~Rechatin, A.~Norlin, A.~Lifschitz, Y.~Glinec, V.~Malka,
  \href{http://dx.doi.org/10.1038/nature05393}{Controlled injection and
  acceleration of electrons in plasma wakefields by colliding laser pulses},
  Nature 444~(7120) (2006) 737--739.
\newblock \href {http://dx.doi.org/10.1038/nature05393}
  {\path{doi:10.1038/nature05393}}.
\newline\urlprefix\url{http://dx.doi.org/10.1038/nature05393}

\bibitem{mangles_monoenergetic_2004}
S.~P.~D. Mangles, C.~D. Murphy, Z.~Najmudin, A.~G.~R. Thomas, J.~L. Collier,
  A.~E. Dangor, E.~J. Divall, P.~S. Foster, J.~G. Gallacher, C.~J. Hooker,
  D.~A. Jaroszynski, A.~J. Langley, W.~B. Mori, P.~A. Norreys, F.~S. Tsung,
  R.~Viskup, B.~R. Walton, K.~Krushelnick,
  \href{http://dx.doi.org/10.1038/nature02939}{Monoenergetic beams of
  relativistic electrons from intense laser–plasma interactions}, Nature
  431~(7008) (2004) 535--538.
\newblock \href {http://dx.doi.org/10.1038/nature02939}
  {\path{doi:10.1038/nature02939}}.
\newline\urlprefix\url{http://dx.doi.org/10.1038/nature02939}

\bibitem{mcguffey_ionization_2010}
C.~McGuffey, A.~G.~R. Thomas, W.~Schumaker, T.~Matsuoka, V.~Chvykov, F.~J.
  Dollar, G.~Kalintchenko, V.~Yanovsky, A.~Maksimchuk, K.~Krushelnick, V.~Y.
  Bychenkov, I.~V. Glazyrin, A.~V. Karpeev,
  \href{http://dx.doi.org/10.1103/PhysRevLett.104.025004}{Ionization {Induced}
  {Trapping} in a {Laser} {Wakefield} {Accelerator}}, Phys. Rev. Lett. 104~(2).
\newblock \href {http://dx.doi.org/10.1103/physrevlett.104.025004}
  {\path{doi:10.1103/physrevlett.104.025004}}.
\newline\urlprefix\url{http://dx.doi.org/10.1103/PhysRevLett.104.025004}

\bibitem{pak_injection_2010}
A.~Pak, K.~A. Marsh, S.~F. Martins, W.~Lu, W.~B. Mori, C.~Joshi,
  \href{http://dx.doi.org/10.1103/PhysRevLett.104.025003}{Injection and
  {Trapping} of {Tunnel}-{Ionized} {Electrons} into {Laser}-{Produced}
  {Wakes}}, Phys. Rev. Lett. 104~(2).
\newblock \href {http://dx.doi.org/10.1103/physrevlett.104.025003}
  {\path{doi:10.1103/physrevlett.104.025003}}.
\newline\urlprefix\url{http://dx.doi.org/10.1103/PhysRevLett.104.025003}

\bibitem{clayton_self-guided_2010}
C.~E. Clayton, J.~E. Ralph, F.~Albert, R.~A. Fonseca, S.~H. Glenzer, C.~Joshi,
  W.~Lu, K.~A. Marsh, S.~F. Martins, W.~B. Mori, A.~Pak, F.~S. Tsung, B.~B.
  Pollock, J.~S. Ross, L.~O. Silva, D.~H. Froula,
  \href{http://dx.doi.org/10.1103/PhysRevLett.105.105003}{Self-{Guided} {Laser}
  {Wakefield} {Acceleration} beyond 1 {GeV} {Using} {Ionization}-{Induced}
  {Injection}}, Phys. Rev. Lett. 105~(10).
\newblock \href {http://dx.doi.org/10.1103/physrevlett.105.105003}
  {\path{doi:10.1103/physrevlett.105.105003}}.
\newline\urlprefix\url{http://dx.doi.org/10.1103/PhysRevLett.105.105003}

\bibitem{faure_laserplasma_2004}
J.~Faure, Y.~Glinec, A.~Pukhov, S.~Kiselev, S.~Gordienko, E.~Lefebvre, J.-P.
  Rousseau, F.~Burgy, V.~Malka,
  \href{http://www.nature.com/nature/journal/v431/n7008/full/nature02963.html}{A
  laser–plasma accelerator producing monoenergetic electron beams}, Nature
  431~(7008) (2004) 541--544.
\newblock \href {http://dx.doi.org/10.1038/nature02963}
  {\path{doi:10.1038/nature02963}}.
\newline\urlprefix\url{http://www.nature.com/nature/journal/v431/n7008/full/nature02963.html}

\bibitem{kalmykov_electron_2011}
S.~Y. Kalmykov, A.~Beck, S.~A. Yi, V.~N. Khudik, M.~C. Downer, E.~Lefebvre,
  B.~A. Shadwick, D.~P. Umstadter,
  \href{http://dx.doi.org/10.1063/1.3566062}{Electron self-injection into an
  evolving plasma bubble: {Quasi}-monoenergetic laser-plasma acceleration in
  the blowout regime}, Physics of Plasmas 18~(5) (2011) 056704.
\newblock \href {http://dx.doi.org/10.1063/1.3566062}
  {\path{doi:10.1063/1.3566062}}.
\newline\urlprefix\url{http://dx.doi.org/10.1063/1.3566062}

\bibitem{liu_all-optical_2011}
J.~S. Liu, C.~Q. Xia, W.~T. Wang, H.~Y. Lu, C.~Wang, A.~H. Deng, W.~T. Li,
  H.~Zhang, X.~Y. Liang, Y.~X. Leng, X.~M. Lu, C.~Wang, J.~Z. Wang,
  K.~Nakajima, R.~X. Li, Z.~Z. Xu,
  \href{http://dx.doi.org/10.1103/PhysRevLett.107.035001}{All-{Optical}
  {Cascaded} {Laser} {Wakefield} {Accelerator} {Using} {Ionization}-{Induced}
  {Injection}}, Phys. Rev. Lett. 107~(3).
\newblock \href {http://dx.doi.org/10.1103/physrevlett.107.035001}
  {\path{doi:10.1103/physrevlett.107.035001}}.
\newline\urlprefix\url{http://dx.doi.org/10.1103/PhysRevLett.107.035001}

\bibitem{pollock_demonstration_2011}
B.~B. Pollock, C.~E. Clayton, J.~E. Ralph, F.~Albert, A.~Davidson, L.~Divol,
  C.~Filip, S.~H. Glenzer, K.~Herpoldt, W.~Lu, K.~A. Marsh, J.~Meinecke, W.~B.
  Mori, A.~Pak, T.~C. Rensink, J.~S. Ross, J.~Shaw, G.~R. Tynan, C.~Joshi,
  D.~H. Froula,
  \href{http://dx.doi.org/10.1103/PhysRevLett.107.045001}{Demonstration of a
  {Narrow} {Energy} {Spread}, \${\textbackslash}textbackslashsim\$ 0.5 {GeV}
  {Electron} {Beam} from a {Two}-{Stage} {Laser} {Wakefield} {Accelerator}},
  Phys. Rev. Lett. 107~(4).
\newblock \href {http://dx.doi.org/10.1103/physrevlett.107.045001}
  {\path{doi:10.1103/physrevlett.107.045001}}.
\newline\urlprefix\url{http://dx.doi.org/10.1103/PhysRevLett.107.045001}

\bibitem{gonsalves_tunable_2011}
A.~J. Gonsalves, K.~Nakamura, C.~Lin, D.~Panasenko, S.~Shiraishi, T.~Sokollik,
  C.~Benedetti, C.~B. Schroeder, C.~G.~R. Geddes, J.~v. Tilborg, J.~Osterhoff,
  E.~Esarey, C.~Toth, W.~P. Leemans,
  \href{http://dx.doi.org/10.1038/nphys2071}{Tunable laser plasma accelerator
  based on longitudinal density tailoring}, Nat Phys 7~(11) (2011) 862--866.
\newblock \href {http://dx.doi.org/10.1038/nphys2071}
  {\path{doi:10.1038/nphys2071}}.
\newline\urlprefix\url{http://dx.doi.org/10.1038/nphys2071}

\bibitem{kim_enhancement_2013}
H.~T. Kim, K.~H. Pae, H.~J. Cha, I.~J. Kim, T.~J. Yu, J.~H. Sung, S.~K. Lee,
  T.~M. Jeong, J.~Lee,
  \href{http://dx.doi.org/10.1103/PhysRevLett.111.165002}{Enhancement of
  {Electron} {Energy} to the {Multi}-{GeV} {Regime} by a {Dual}-{Stage}
  {Laser}-{Wakefield} {Accelerator} {Pumped} by {Petawatt} {Laser} {Pulses}},
  Phys. Rev. Lett. 111~(16).
\newblock \href {http://dx.doi.org/10.1103/physrevlett.111.165002}
  {\path{doi:10.1103/physrevlett.111.165002}}.
\newline\urlprefix\url{http://dx.doi.org/10.1103/PhysRevLett.111.165002}

\bibitem{wang_quasi-monoenergetic_2013}
X.~Wang, R.~Zgadzaj, N.~Fazel, Z.~Li, S.~A. Yi, X.~Zhang, W.~Henderson, Y.-Y.
  Chang, R.~Korzekwa, H.-E. Tsai, C.-H. Pai, H.~Quevedo, G.~Dyer, E.~Gaul,
  M.~Martinez, A.~C. Bernstein, T.~Borger, M.~Spinks, M.~Donovan, V.~Khudik,
  G.~Shvets, T.~Ditmire, M.~C. Downer,
  \href{http://www.nature.com/ncomms/2013/130611/ncomms2988/full/ncomms2988.html}{Quasi-monoenergetic
  laser-plasma acceleration of electrons to 2 {GeV}}, Nature Communications 4
  (2013) 1988.
\newblock \href {http://dx.doi.org/10.1038/ncomms2988}
  {\path{doi:10.1038/ncomms2988}}.
\newline\urlprefix\url{http://www.nature.com/ncomms/2013/130611/ncomms2988/full/ncomms2988.html}

\bibitem{golovin_tunable_2015}
G.~Golovin, S.~Chen, N.~Powers, C.~Liu, S.~Banerjee, J.~Zhang, M.~Zeng,
  Z.~Sheng, D.~Umstadter,
  \href{http://dx.doi.org/10.1103/PhysRevSTAB.18.011301}{Tunable monoenergetic
  electron beams from independently controllable laser-wakefield acceleration
  and injection}, Phys. Rev. ST Accel. Beams 18~(1).
\newblock \href {http://dx.doi.org/10.1103/physrevstab.18.011301}
  {\path{doi:10.1103/physrevstab.18.011301}}.
\newline\urlprefix\url{http://dx.doi.org/10.1103/PhysRevSTAB.18.011301}

\bibitem{lee_optimization_2017}
P.~Lee, G.~Maynard, T.~L. Audet, R.~Lehe, J.~L. Vay, B.~Cros,
  \href{http://arxiv.org/abs/1711.01613}{Optimization of laser-plasma injector
  via beam loading effects using ionization-induced injection},
  arXiv:1711.01613 [physics]ArXiv: 1711.01613.
\newline\urlprefix\url{http://arxiv.org/abs/1711.01613}

\bibitem{audet_electron_2016}
T.~L. Audet, F.~G. Desforges, A.~Maitrallain, S.~D. Dufrénoy, M.~Bougeard,
  G.~Maynard, P.~Lee, M.~Hansson, B.~Aurand, A.~Persson, I.~G. González,
  P.~Monot, C.-G. Wahlström, O.~Lundh, B.~Cros,
  \href{http://dx.doi.org/10.1016/j.nima.2016.01.035}{Electron injector for
  compact staged high energy accelerator}, Nuclear Instruments and Methods in
  Physics Research Section A: Accelerators, Spectrometers, Detectors and
  Associated Equipment\href {http://dx.doi.org/10.1016/j.nima.2016.01.035}
  {\path{doi:10.1016/j.nima.2016.01.035}}.
\newline\urlprefix\url{http://dx.doi.org/10.1016/j.nima.2016.01.035}

\bibitem{weller_tensorial_1998}
H.~G. Weller, G.~Tabor, H.~Jasak, C.~Fureby,
  \href{http://dx.doi.org/10.1063/1.168744}{A tensorial approach to
  computational continuum mechanics using object-oriented techniques},
  Computers in Physics 12~(6) (1998) 620.
\newblock \href {http://dx.doi.org/10.1063/1.168744}
  {\path{doi:10.1063/1.168744}}.
\newline\urlprefix\url{http://dx.doi.org/10.1063/1.168744}

\bibitem{audet_gas_nodate}
T.~L. Audet, P.~Lee, G.~Maynard, S.~Dobosz~Dufrénoy, A.~Maitrallain,
  M.~Bougeard, P.~Monot, B.~Cros, Gas cell density characterization for laser
  wakefield acceleration, Nuclear Instruments and Methods in Physics Research
  Section A: Accelerators, Spectrometers, Detectors and Associated
  Equipment~(this issue).

\bibitem{lee_dynamics_2016}
P.~Lee, G.~Maynard, T.~L. Audet, B.~Cros, R.~Lehe, J.-L. Vay,
  \href{http://link.aps.org/doi/10.1103/PhysRevAccelBeams.19.112802}{Dynamics
  of electron injection and acceleration driven by laser wakefield in tailored
  density profiles}, Physical Review Accelerators and Beams 19~(11) (2016)
  112802.
\newblock \href {http://dx.doi.org/10.1103/PhysRevAccelBeams.19.112802}
  {\path{doi:10.1103/PhysRevAccelBeams.19.112802}}.
\newline\urlprefix\url{http://link.aps.org/doi/10.1103/PhysRevAccelBeams.19.112802}

\bibitem{vay_novel_2012}
J.-L. Vay, D.~P. Grote, R.~H. Cohen, A.~Friedman,
  \href{http://dx.doi.org/10.1088/1749-4699/5/1/014019}{Novel methods in the
  {Particle}-{In}-{Cell} accelerator {Code}-{Framework} {Warp}}, Comput. Sci.
  Disc. 5~(1) (2012) 014019.
\newblock \href {http://dx.doi.org/10.1088/1749-4699/5/1/014019}
  {\path{doi:10.1088/1749-4699/5/1/014019}}.
\newline\urlprefix\url{http://dx.doi.org/10.1088/1749-4699/5/1/014019}

\bibitem{lifschitz_particle--cell_2009}
A.~F. Lifschitz, X.~Davoine, E.~Lefebvre, J.~Faure, C.~Rechatin, V.~Malka,
  \href{http://dx.doi.org/10.1016/j.jcp.2008.11.017}{Particle-in-{Cell}
  modelling of laser–plasma interaction using {Fourier} decomposition},
  Journal of Computational Physics 228~(5) (2009) 1803--1814.
\newblock \href {http://dx.doi.org/10.1016/j.jcp.2008.11.017}
  {\path{doi:10.1016/j.jcp.2008.11.017}}.
\newline\urlprefix\url{http://dx.doi.org/10.1016/j.jcp.2008.11.017}

\bibitem{davidson_implementation_2015}
A.~Davidson, A.~Tableman, W.~An, F.~S. Tsung, W.~Lu, J.~Vieira, R.~A. Fonseca,
  L.~O. Silva, W.~B. Mori,
  \href{http://dx.doi.org/10.1016/j.jcp.2014.10.064}{Implementation of a hybrid
  particle code with a {PIC} description in r-z and a gridless description in
  phi into {OSIRIS}}, Journal of Computational Physics 281 (2015) 1063--1077.
\newblock \href {http://dx.doi.org/10.1016/j.jcp.2014.10.064}
  {\path{doi:10.1016/j.jcp.2014.10.064}}.
\newline\urlprefix\url{http://dx.doi.org/10.1016/j.jcp.2014.10.064}

\bibitem{lee_modeling_2015}
P.~Lee, T.~L. Audet, R.~Lehe, J.-L. Vay, G.~Maynard, B.~Cros,
  \href{http://dx.doi.org/10.1016/j.nima.2015.12.036}{Modeling laser-driven
  electron acceleration using {WARP} with {Fourier} decomposition}, Nuclear
  Instruments and Methods in Physics Research Section A: Accelerators,
  Spectrometers, Detectors and Associated Equipment\href
  {http://dx.doi.org/10.1016/j.nima.2015.12.036}
  {\path{doi:10.1016/j.nima.2015.12.036}}.
\newline\urlprefix\url{http://dx.doi.org/10.1016/j.nima.2015.12.036}

\bibitem{desforges_dynamics_2014}
F.~G. Desforges, B.~S. Paradkar, M.~Hansson, J.~Ju, L.~Senje, T.~L. Audet,
  A.~Persson, S.~Dobosz-Dufrénoy, O.~Lundh, G.~Maynard, P.~Monot, J.-L. Vay,
  C.-G. Wahlström, B.~Cros,
  \href{http://dx.doi.org/10.1063/1.4903845}{Dynamics of ionization-induced
  electron injection in the high density regime of laser wakefield
  acceleration}, Physics of Plasmas 21~(12) (2014) 120703.
\newblock \href {http://dx.doi.org/10.1063/1.4903845}
  {\path{doi:10.1063/1.4903845}}.
\newline\urlprefix\url{http://dx.doi.org/10.1063/1.4903845}

\bibitem{ammosov_tunnel_1986}
M.~Ammosov, N.~Delone, V.~Krainov, Tunnel ionization of complex atoms and of
  atomic ions in an alternating electric field, Sov. Phys. JETP 64 (1986) 1191.

\bibitem{barber_measured_2017}
S.~Barber, J.~van Tilborg, C.~Schroeder, R.~Lehe, H.-E. Tsai, K.~Swanson,
  S.~Steinke, K.~Nakamura, C.~Geddes, C.~Benedetti, E.~Esarey, W.~Leemans,
  \href{https://link.aps.org/doi/10.1103/PhysRevLett.119.104801}{Measured
  {Emittance} {Dependence} on the {Injection} {Method} in {Laser} {Plasma}
  {Accelerators}}, Physical Review Letters 119~(10) (2017) 104801.
\newblock \href {http://dx.doi.org/10.1103/PhysRevLett.119.104801}
  {\path{doi:10.1103/PhysRevLett.119.104801}}.
\newline\urlprefix\url{https://link.aps.org/doi/10.1103/PhysRevLett.119.104801}

\end{thebibliography}

\end{document}